\documentclass[conference]{IEEEtran}
\IEEEoverridecommandlockouts
\usepackage{cite}
\usepackage{amsmath,amssymb,amsfonts}
\usepackage{algorithmic}
\usepackage{graphicx}
\usepackage{textcomp}
\usepackage{hyperref}
\usepackage{xcolor}
\usepackage{url}
\usepackage{booktabs}
\usepackage[acronym, nomain, nopostdot]{glossaries}
\usepackage{color}
\usepackage{import}
\usepackage{tikz, pgfplots, pgfkeys}
\usetikzlibrary{arrows.meta}
\usepackage{standalone}
\def\BibTeX{{\rm B\kern-.05em{\sc i\kern-.025em b}\kern-.08em
    T\kern-.1667em\lower.7ex\hbox{E}\kern-.125emX}}

\hypersetup{nolinks=true}

\newacronym{ADRENALINE}{ADRENALINE}{\emph{\underline{A}ttention-based \underline{D}eep \underline{RE}current \underline{N}eural-network for loc\underline{ALI}zing sou\underline{N}d \underline{E}vents}}
\newacronym{CNN}{CNN}{convolutional neural network}
\newacronym{DoA}{DoA}{direction-of-arrival}
\newacronym{GMM}{GMM}{Gaussian mixture model}
\newacronym{GRU}{GRU}{gated recurrent unit}
\newacronym{LSTM}{LSTM}{long-short term memory}
\newacronym{MSE}{MSE}{mean squared error}
\newacronym{MUSIC}{MUSIC}{multiple signal classification}
\newacronym{NLP}{NLP}{natural language processing}
\newacronym{ReLU}{ReLU}{rectified linear unit}
\newacronym{RNN}{RNN}{recurrent neural network}
\newacronym{SEL}{SEL}{sound event localization}
\newacronym{SELD}{SELD}{sound event localization and detection}
\newacronym{SRP-PHAT}{SRP-PHAT}{steered  response  power  phase  transform}
\newacronym{STFT}{STFT}{short-time Fourier transform}

\begin{document}

\title{Exploiting Attention-based Sequence-to-Sequence Architectures for Sound Event Localization\\
}

\author{\IEEEauthorblockN{
		Christopher Schymura\IEEEauthorrefmark{1}, 
		Tsubasa Ochiai\IEEEauthorrefmark{2},
		Marc Delcroix\IEEEauthorrefmark{2},
		Keisuke Kinoshita\IEEEauthorrefmark{2},
		Tomohiro Nakatani\IEEEauthorrefmark{2},\\
		Shoko Araki\IEEEauthorrefmark{2},
		Dorothea Kolossa\IEEEauthorrefmark{1}
	}
	\IEEEauthorblockA{\IEEEauthorrefmark{1}
		Institute of Communication Acoustics\\
		Ruhr University Bochum, Bochum, Germany\\
		Email: christopher.schymura@rub.de}
	\IEEEauthorblockA{\IEEEauthorrefmark{2}
		NTT Communication Science Laboratories\\
		NTT Corporation, Kyoto, Japan}}

\maketitle

\begin{abstract}
Sound event localization frameworks based on deep neural networks have shown increased robustness with respect to reverberation and noise in comparison to classical parametric approaches. In particular, recurrent architectures that incorporate temporal context into the estimation process seem to be well-suited for this task. This paper proposes a novel approach to sound event localization by utilizing an attention-based sequence-to-sequence model. These types of models have been successfully applied to problems in natural language processing and automatic speech recognition. In this work, a multi-channel audio signal is encoded to a latent representation, which is subsequently decoded to a sequence of estimated directions-of-arrival. Herein, attentions allow for capturing temporal dependencies in the audio signal by focusing on specific frames that are relevant for estimating the activity and direction-of-arrival of sound events at the current time-step. The framework is evaluated on three publicly available datasets for sound event localization. It yields superior localization performance compared to state-of-the-art methods in both anechoic and reverberant conditions.
\end{abstract}


\section{Introduction}
\label{sec:introduction}
\Gls{SEL} considers estimating the spatial positions of sound events from audio signals. The definition of a sound event refers to a sound produced by a variety of sources, e.g. speech or a musical instrument, which oftentimes is a time-varying quantity with respect to volume and position. Many applications rely on accurate \gls{SEL}, as it is an important processing step for speech enhancement in hearing aids~\cite{Li2018}, acoustic monitoring for industrial applications~\cite{Grobler2017}, robotics~\cite{Evers2018} and many others. The importance of localization and tracking algorithms for audio signal processing has also been stressed during the recent LOCATA challenge~\cite{Lollmann2018}, where a wide range of contributions tackled this problem in a diverse set of challenging scenarios, cf.~\cite{Evers2019} for details.

However, many classical frameworks for \gls{SEL} are based on parametric approaches, e.g. the \gls{SRP-PHAT}~\cite{DiBiase2001} or \gls{MUSIC}~\cite{Schmidt1986} methods. Furthermore, approaches using learning-based techniques like logistic regression~\cite{Trowitzsch2020} or independent component analysis~\cite{Sawada2005} have also been proposed.

In addition, approaches utilizing deep neural networks for \gls{SEL} have been developed recently. For instance, the method described in~\cite{Chakrabarty2017} proposed a \gls{CNN} operating on the phase of the audio input in the \gls{STFT} domain to predict the \gls{DoA} of a sound event. The work of He et al.~\cite{He2018} introduced a similar framework for robotics applications, which used features derived from \gls{SRP-PHAT} as input to a \gls{CNN}-based \gls{SEL} system. This approach was further developed in~\cite{Adavanne2018b}, where the complex \gls{STFT} spectrum was used as input to a \gls{CNN}, followed by a \gls{RNN} based on \glspl{GRU}. A spatial pseudo spectrum, similar to classical parametric approaches like \gls{MUSIC}, was obtained from the recurrent output of the network. This representation was used to obtain \gls{DoA} estimates for each sound event. Further extensions towards \gls{SELD} were proposed in~\cite{Adavanne2018a}, which considered a joint estimation of sound type and spatial location. This system was recently adapted to tackle dynamic scenarios with moving sound sources~\cite{Adavanne2019}. Models that exploit temporal context for \gls{SEL} have shown superior performance compared to conventional feed-forward networks~\cite{Adavanne2018a, Wang2018, Adavanne2019}. 

This paper proposes a recurrent architecture for \gls{SEL} in scenarios with multiple, potentially overlapping sound sources, via an attention-based sequence-to-sequence approach to handle temporal dependencies. Such models were initially proposed in the context of neural machine translation~\cite{Bahdanau2014}, but have also been utilized for other tasks like automatic speech recognition~\cite{Tuske2019}. Compared to a conventional recurrent network architecture, the attention mechanism facilitates an improved capturing of the temporal structure in the input sequence. This is achieved by exclusively focusing on parts of the audio signal that are relevant for predicting the spatial source locations at the current frame. This also enables the model to account for long-term dependencies when localizing sound events, which is generally not possible with classical methods that rely on the first-order Markov assumption~\cite{Chakrabarty2014}.
\section{System overview}
\label{sec:system_overview}
\begin{figure*}[t]
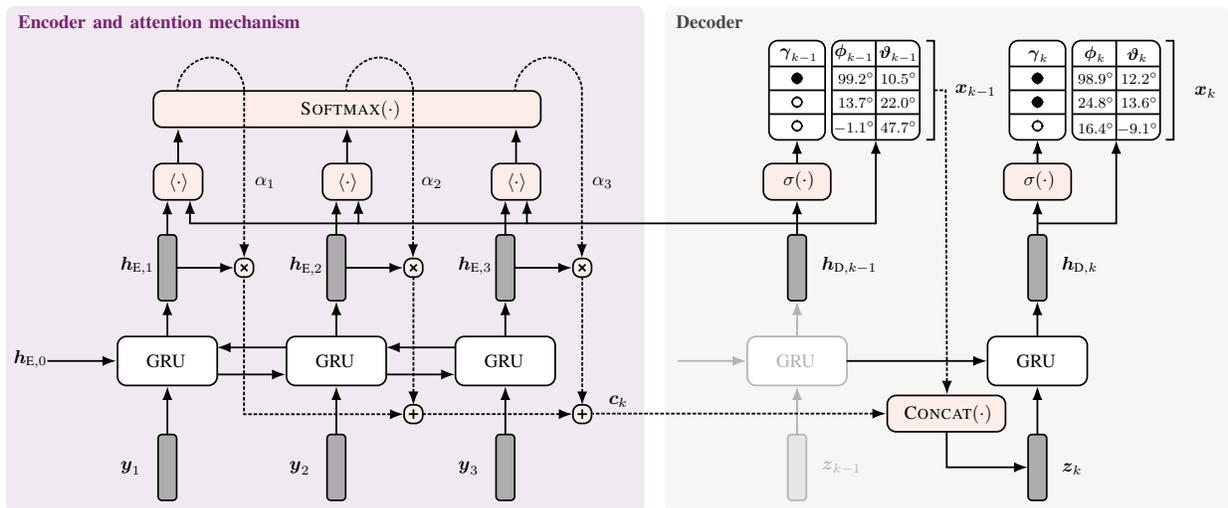

	\centering
	\hspace{-1.0cm}
	\includestandalone[width=0.95\textwidth]{figs/model_architecture}
	\caption{General overview of the {ADRENALINE} model architecture. The box on the left shows an exemplary encoding process for three discrete time-steps from~\(k^{\prime}=1,\,\ldots,\,3\), where~\(\boldsymbol{y}_{k^{\prime}}\) denotes the observed input features and \(\boldsymbol{h}_{\text{E}, k^{\prime}}\) corresponds to the hidden state of the encoder {GRUs}. The attention weights \(\alpha_{k^{\prime}}\) are computed via scaled dot products between the encoder hidden states and the corresponding decoder hidden state \(\boldsymbol{h}_{\text{D}, k-1}\) from the previous decoding time-step. A context vector \(\boldsymbol{c}_{k}\) is derived as a weighted sum of the encoder hidden states, using the attention weights. The output of the decoder \(\boldsymbol{x}_{k}\) is composed of the source activity indicator \(\boldsymbol{\gamma}_{k}\) and the corresponding source {DoAs}, comprising azimuth \(\boldsymbol{\phi}_{k}\) and elevation \(\boldsymbol{\vartheta}_{k}\). A concatenation of the decoder output from the previous time-step and the current context vector serves as input to the decoder, as shown in the box on the right.}
	\label{fig:model_architecture}
\end{figure*}

The proposed \gls{SEL} model is based on a recurrent encoder-decoder architecture with attentions~\cite{Bahdanau2014}. Therefore, it is termed \gls{ADRENALINE} in this work, to stress the underlying design for the desired application. The model essentially consists of three building-blocks: a \gls{CNN}-based feature extraction, the encoder \gls{RNN} and the decoder \gls{RNN} with an attention mechanism, which are depicted in Fig.~\ref{fig:model_architecture}. These components will be described in the following.

\subsection{Feature extraction}
\label{subsec:feature_extraction}
To extract input features from a spectral representation of acquired audio signals, the \gls{CNN}-based feature extractor from SELDNet~\cite{Adavanne2018a} was utilized in this work. The \gls{ADRENALINE} architecture operates on fixed-length chunks of audio data: Given an acoustic waveform with \(C\) channels, it is divided into non-overlapping chunks of \(500\,\text{ms}\) duration. An \gls{STFT}, using a 2048-point Hamming window with a frame length of \(40\,\text{ms}\) and \(20\,\text{ms}\) shift, is performed within each chunk. The complex spectrum is separated into its magnitude and phase components, which are concatenated to an input tensor of dimension \(K \times L \times 2C\), where \(K = 25\) is the number of frames within one chunk and \(L = 1024\) is the number of non-redundant bins in the complex spectrum.

The input tensor is fed into a \gls{CNN} with three layers. Each layer is composed of \(64\) filters with kernel size \(3 \times 3\), followed by batch normalization, \glspl{ReLU} and max-pooling with kernel size \(1 \times 8\) in the first two layers and \(1 \times 2\) in the third layer. The output of the third layer is reshaped to yield \(K \times D_{y}\) feature sequences, where \(D_{y} = 512\) is the feature dimension. The elements of the feature sequence will be denoted as \(\boldsymbol{y}_{k^{\prime}} \in \mathbb{R}^{D_{y}}\), where \(k^{\prime} = 1,\,\ldots,\,K\) represent the encoder time indices.

\subsection{Encoder}
\label{subsec:encoder}
A conventional \gls{RNN}-based encoder architecture is used in the \gls{ADRENALINE} framework. The input feature sequence is fed into a bidirectional \gls{RNN}, which yields a hidden state representation \(\boldsymbol{h}_{\text{E}, k^{\prime}} \in \mathbb{R}^{2 D_{h}}\) at each time step, according to
\begin{equation}
\boldsymbol{h}_{\text{E}, k^{\prime}} = f_{\text{E}}(\boldsymbol{h}_{\text{E}, k^{\prime}-1},\,\boldsymbol{y}_{k^{\prime}}).
\label{eqn:encoder}
\end{equation}
Due to the bidirectional implementation of the \gls{RNN}, the hidden state is composed of the states obtained during the forward and backward passes~\cite{Schuster1997}, which is not explicitly shown in Eq.~\eqref{eqn:encoder}, cf.~\cite{Bahdanau2014} for details. A \gls{GRU} with a hidden dimension of \(D_{h} = 64\) is used as the underlying \gls{RNN} cell. It is initialized with zeros for \(k^{\prime} = 0\).

\subsection{Decoder and attention mechanism}
\label{subsec:decoder}
As shown in Fig.~\ref{fig:model_architecture}, at each decoder time index \(k\), the decoder takes as input a vector \(\boldsymbol{z}_{k}\), which is composed of a decoder output vector \(\boldsymbol{x}_{k-1}\) from the previous time-step and a context vector~\(\boldsymbol{c}_{k}\), generated by the attention mechanism. The context vector is computed according to
\begin{equation}
\boldsymbol{c}_{k} = \sum_{k^{\prime} = 1}^{K} \alpha_{k^{\prime}} \boldsymbol{h}_{\text{E}, k^{\prime}},\label{eqn:context_vector}
\end{equation}
where \(\alpha_{k}\) denote the attention weights. Each attention weight is obtained via a scaled dot-product~\cite{Vaswani2017} between the encoder hidden state and the corresponding previous decoder hidden state, followed by a softmax activation function
\begin{equation}
\alpha_{k^{\prime}} = \textsc{Softmax}\Big(\frac{1}{\sqrt{2D_{h}}}\boldsymbol{h}_{\text{E}, k^{\prime}}^{\mathrm{T}} \boldsymbol{h}_{\text{D}, k-1}\Big).
\label{eqn:attention_weight}
\end{equation}
The decoder output \(\boldsymbol{x}_{k}\) represents the current information about detected sound events and their locations. It is composed of three elements: a vector~\(\boldsymbol{\gamma}_{k} \in \mathbb{R}^{S}\) indicating source activity and vectors of corresponding azimuth and elevation angles, \(\boldsymbol{\phi}_{k} \in \mathbb{R}^{S}\) and~\(\boldsymbol{\vartheta}_{k} \in \mathbb{R}^{S}\), respectively. Each \gls{DoA} corresponds to the source activity indicator in the same row of the source activity vector. Their dimension \(S\) indicates the maximum number of sources that the framework can handle simultaneously. It is set to \(S = 4\) in this work. This representation was inspired by the SELDNet architecture~\cite{Adavanne2018a}, which utilized a similar output structure, but additionally performed sound event detection, which is omitted in the framework proposed here. The joint decoder output \(\boldsymbol{x}_{k} \in \mathbb{R}^{3S}\) is obtained by stacking the individual output elements into a single vector. Hence, by utilizing the result from Eq.~\eqref{eqn:context_vector}, the decoder input vector can be represented as \(\boldsymbol{z}_{k} = \begin{bmatrix}\boldsymbol{c}_{k} & \boldsymbol{x}_{k - 1}\end{bmatrix}^{\mathrm{T}} \in \mathbb{R}^{2D_{h} + 3S}\).

The decoding step for estimating source activities and locations can now be expressed via the decoder \gls{RNN}
\begin{equation}
\boldsymbol{h}_{\text{D}, k} = f_{\text{D}}(\boldsymbol{h}_{\text{D}, k-1},\,\boldsymbol{z}_{k}),
\label{eqn:decoder}
\end{equation}
whose outputs are \(\boldsymbol{\gamma}_{k} = \sigma(\boldsymbol{W}^{\boldsymbol{\gamma}} \boldsymbol{h}_{\text{D}, k})\), \(\boldsymbol{\phi}_{k} = \boldsymbol{W}^{\boldsymbol{\phi}} \boldsymbol{h}_{\text{D}, k}\) and \(\boldsymbol{\vartheta}_{k} = \boldsymbol{W}^{\boldsymbol{\vartheta}} \boldsymbol{h}_{\text{D}, k}\), with weights \(\boldsymbol{W}^{\boldsymbol{\gamma}}\), \(\boldsymbol{W}^{\boldsymbol{\phi}}\) and \(\boldsymbol{W}^{\boldsymbol{\vartheta}}\), where \(\sigma(\cdot)\) denotes the sigmoid nonlinearity.

\subsection{Loss function}
\label{subsec:loss_function}
The decoder output representation introduced in the previous section requires a form of loss that is tailored to the \gls{SEL} task. Therefore, a loss function with two components
\begin{equation}
\mathcal{L}_{\text{SEL}} = \mathcal{L}_{\text{ACT}}(\hat{\boldsymbol{\gamma}},\,\boldsymbol{\gamma}) + \lambda \mathcal{L}_{\text{DOA}}(\hat{\boldsymbol{\phi}},\,\hat{\boldsymbol{\vartheta}},\,\boldsymbol{\phi},\,\boldsymbol{\vartheta},\,\boldsymbol{\gamma})
\label{eqn:loss_function}
\end{equation}
is proposed, where \(\lambda \in \mathbb{R}_{+}\) is a scaling factor. The discrete decoder time index \(k\) is omitted here for notational convenience. Eq.~\eqref{eqn:loss_function} incorporates a binary cross-entropy loss term \(\mathcal{L}_{\text{ACT}}(\hat{\boldsymbol{\gamma}},\,\boldsymbol{\gamma})\)~\cite[Chap.~5]{Goodfellow2016} evaluating the difference between estimated and ground-truth source activities~\(\hat{\boldsymbol{\gamma}}\) and~\(\boldsymbol{\gamma}\), respectively. The second term is based on the \gls{DoA} error~\cite{Adavanne2018b}
\begin{equation}
\boldsymbol{\xi} = \arccos\Big(\sin(\hat{\boldsymbol{\phi}}) \sin(\boldsymbol{\phi}) + \cos(\hat{\boldsymbol{\phi}}) \cos(\boldsymbol{\phi}) \cos(\boldsymbol{\vartheta} - \hat{\boldsymbol{\vartheta}})\Big),
\label{eqn:doa_error}
\end{equation}
which measures the angle between the estimated azimuth \(\hat{\boldsymbol{\phi}}\) and elevation \(\hat{\boldsymbol{\vartheta}}\) and the ground-truth \gls{DoA}, given by \(\boldsymbol{\phi}\) and \(\boldsymbol{\vartheta}\). The trigonometric functions in Eq.~\eqref{eqn:doa_error} are applied element-wise, so that the resulting vector~\(\boldsymbol{\xi}\) contains a single \gls{DoA} error for each potential source. The \gls{DoA} error showed slightly better performance when dealing with angular values, compared to the more commonly used \gls{MSE} loss.

As only active sources shall be considered in the \gls{DoA} loss, the \gls{DoA} errors of non-active sources are omitted by incorporating the ground-truth source activity vector as
\begin{equation}
\mathcal{L}_{\text{DOA}}(\hat{\boldsymbol{\phi}},\,\hat{\boldsymbol{\vartheta}},\,\boldsymbol{\phi},\,\boldsymbol{\vartheta},\,\boldsymbol{\gamma}) = \frac{1}{S} \boldsymbol{\xi}^{\mathrm{T}} \boldsymbol{\gamma},
\label{eqn:doa_loss}
\end{equation}
which effectively blocks the gradient flow for non-active sources during back-propagation. 

It should be noted that the \gls{DoA} loss presented in Eq.~\eqref{eqn:doa_loss} does not take potentially occurring source permutations between consecutive time-steps into account. Therefore, the \gls{DoA} error given in Eq.~\eqref{eqn:doa_error} is computed for all possible source permutations during the calculation of the loss. This is computationally feasible, as the maximum number of sources set to \(S = 4\) is small and all computations can be run in parallel. The lowest resulting \gls{DoA} error is then selected for evaluating Eq.~\eqref{eqn:doa_loss}. Similar approaches have also been used in other application domains, e.g. for speech separation in~\cite{Yu2017}.

\section{Evaluation}
\label{sec:evaluation}
The proposed \gls{ADRENALINE} framework is evaluated on three publicly available datasets and compared to related \gls{SEL} baseline methods. The program code is accessible online\footnote{\url{https://github.com/rub-ksv/adrenaline}} and the detailed training procedure and experimental setup is described in the following.

\subsection{Datasets}
\label{subsec:datasets}
The ANSYN, RESYN and REAL datasets introduced in~\cite{Adavanne2018a} were selected as suitable evaluation corpora in this study. These datasets provide a number of first order Ambisonic format recordings that is large enough to train deep learning models for \gls{SEL}.

The first two were synthesized using simulated anechoic and reverberant impulse responses, whereas the latter utilized impulse responses recorded in real environments. All datasets are assembled with a similar structure: they comprise three subsets each, corresponding to either no, at most two, or at most three temporally overlapping sources. Each subset provides three cross-validation splits with 300 audio files each. These files are further divided into 240 files for training and 60 for validation. All audio files are sampled with \(44.1\,\text{kHz}\), have a duration of \(30\,\text{s}\) and are encoded in the Ambisonics format. The audio signals in the synthetic datasets were created using sound events from 11 different classes, covering the full azimuth range and an elevation range from \(-60^{\circ}\) to \(60^{\circ}\). For the recorded impulse response dataset, sound events from 8 classes were used for synthesizing the audio signals. Herein, again the full azimuth range was covered and the elevation range was restricted between \(-40^{\circ}\) and \(40^{\circ}\).

\subsection{Baseline methods}
\label{subsec:baseline_methods}
Two baseline methods were employed in this work to compare the performance of the proposed model to similar \gls{SEL} frameworks. The first baseline model builds upon the \gls{CNN} feature extraction stage described in Sec.~\ref{subsec:feature_extraction}. It simply adds a fully connected layer, which outputs source activity and \gls{DoA} vectors. This is a feed-forward model with no recurrent elements and will be referred to as CNN in the experiments. 

Additionally, a variant of the SELDNet architecture~\cite{Adavanne2018a} is utilized as a second baseline. As the name suggests, it was initially proposed for \gls{SELD} tasks, predicting location and class of a sound event, where it achieved state-of-the-art performance. It shares the same \gls{CNN}-based feature extraction stage with the proposed \gls{ADRENALINE} framework and is also based on bidirectional \glspl{GRU}. The main difference between the methods is that SELDNet does not incorporate any attention mechanisms. Instead, it directly utilizes the output of the recurrent layers for estimation.

To enable a fair comparison to the proposed \gls{ADRENALINE} framework, the SELDNet model is modified here by removing information about the sound class from the output layer. This essentially yields a system output identical to the one proposed in Sec.~\ref{subsec:decoder}. The model is referred to as SELDNet(m) in the following, denoting the modified output.

\subsection{Performance metrics}
\label{subsec:metrics}
Frame recall and \gls{DoA} error are used as metrics for evaluating \gls{SEL} performance. The frame recall metric~\cite{Adavanne2018a} describes the percentage of frames, where the estimated number of active sources matches the ground-truth. During evaluation, a source is considered active if the corresponding element in the source activity vector exceeds a threshold of \(0.5\). The \gls{DoA} error is computed as described in Eq.~\eqref{eqn:doa_error}, where only active sources are considered. The Hungarian algorithm~\cite{Kuhn1955} is used to solve the assignment problem between multiple estimated \glspl{DoA} and the ground-truth, which yields the smallest possible \gls{DoA} error.

\subsection{Experimental setup}
\label{subsec:experimental_setup}
All models were implemented in PyTorch~\cite{Paszke2019}. The baseline model parameterization was chosen as reported in~\cite{Adavanne2018a}. The parameters of the proposed \gls{ADRENALINE} model were specified as described in Sec.~\ref{sec:system_overview}. The \gls{SEL} loss function given in Eq.~\eqref{eqn:loss_function} was used with a scaling factor of \(\lambda=1\) for optimizing all evaluated models. All model parameters were initialized using the Kaiming initialization method from~\cite{He2015}. Training was conducted using the AdamW optimizer~\cite{Loshchilov2019} with batch-size~\(256\) and a learning rate of \(0.0002\). The learning rate was varied by adopting the scheduling scheme proposed in~\cite{Vaswani2017}, which resulted in an increased training stability. Separate models were trained for each cross-validation fold, utilizing early stopping with a maximum number of \(200\) epochs.
\begin{table}[t]
	\caption{Frame recall in percent grouped by dataset. The metric is averaged over all subsets and cross-validation folds.}
	\centering
	\renewcommand{\arraystretch}{1.33}
	\footnotesize
	\begin{tabular}{lccc}
		\toprule
		& ANSYN & RESYN & REAL \\
		\midrule
		CNN & \textbf{87.48} & 71.91 & 72.07 \\
		SELDNet(m) & 85.78 & \textbf{72.46} & 69.63 \\
		ADRENALINE & 84.83 & 71.18 & \textbf{72.08} \\
		\bottomrule
	\end{tabular}
	\label{tab:frame_recall}
\end{table}

\begin{figure}[t]
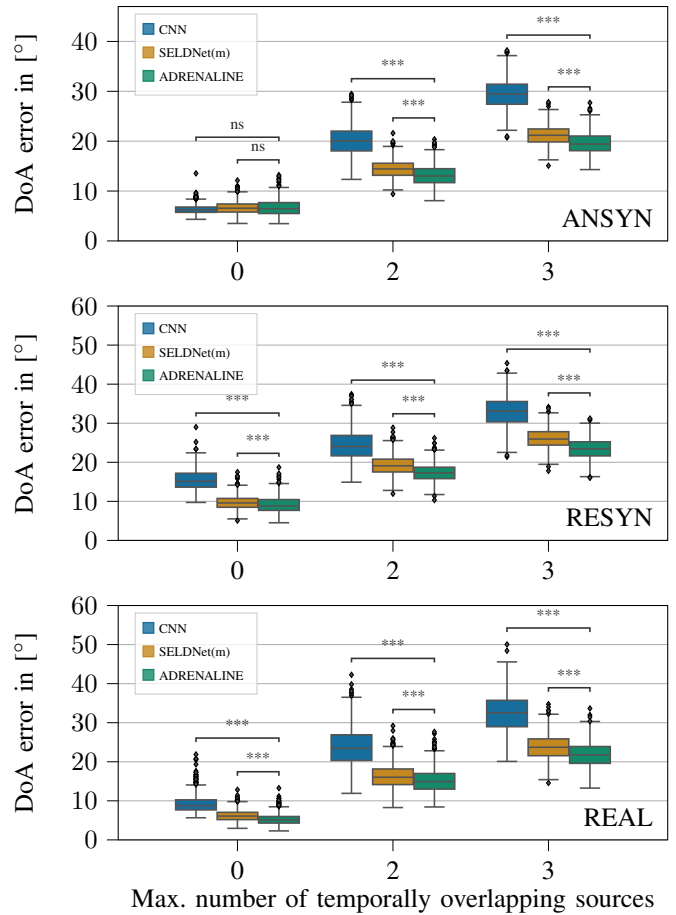

	\centering
	\input{figs/doa_error_ansim}%
	\input{figs/doa_error_resim}%
	\input{figs/doa_error_real}%
	\caption{\gls{DoA} errors obtained in all cross-validation folds for all three datasets, shown as grouped box-plots. Asterisks indicate a statistically significant difference, where \({\ast\ast}\ast\) denotes \(p < 0.001\).}
	\label{fig:doa_errors}
\end{figure}

\section{Results and discussion}
\label{sec:results}
The experimental results are summarized in Tab.~\ref{tab:frame_recall} and Fig.~\ref{fig:doa_errors}. The Mann-Whitney-\(U\) test~\cite{Mann1947} was used to show statistically significant differences in \gls{DoA} error between the proposed \gls{ADRENALINE} framework and the baselines. This nonparametric test was chosen, because the application of a Shapiro-Wilk test~\cite{Shapiro1965} on the resulting \gls{DoA} errors indicated that the individual samples are not normally distributed.

The resulting \gls{DoA} errors indicate that the proposed \gls{ADRENALINE} framework outperforms both baseline methods on all three datasets, where the performance differences are statistically significant in most cases. This shows that employing a sequence-to-sequence architecture for \gls{SEL} yields improved localization performance compared to standard recurrent architectures and feed-forward networks in anechoic as well as reverberant environments. The performance benefit of the \gls{ADRENALINE} framework is especially prominent in scenarios with multiple overlapping sources. 

\begin{figure}[t]
	\centering
\begin{tikzpicture}

\begin{axis}[
colorbar style={ylabel={}},
colormap={mymap}{[1pt]
  rgb(0pt)=(0.1178,0,0);
  rgb(1pt)=(0.195857,0.102869,0.102869);
  rgb(2pt)=(0.250661,0.145479,0.145479);
  rgb(3pt)=(0.295468,0.178174,0.178174);
  rgb(4pt)=(0.334324,0.205738,0.205738);
  rgb(5pt)=(0.369112,0.230022,0.230022);
  rgb(6pt)=(0.400892,0.251976,0.251976);
  rgb(7pt)=(0.430331,0.272166,0.272166);
  rgb(8pt)=(0.457882,0.290957,0.290957);
  rgb(9pt)=(0.483867,0.308607,0.308607);
  rgb(10pt)=(0.508525,0.3253,0.3253);
  rgb(11pt)=(0.532042,0.341178,0.341178);
  rgb(12pt)=(0.554563,0.356348,0.356348);
  rgb(13pt)=(0.576204,0.370899,0.370899);
  rgb(14pt)=(0.597061,0.3849,0.3849);
  rgb(15pt)=(0.617213,0.39841,0.39841);
  rgb(16pt)=(0.636729,0.411476,0.411476);
  rgb(17pt)=(0.655663,0.424139,0.424139);
  rgb(18pt)=(0.674066,0.436436,0.436436);
  rgb(19pt)=(0.69198,0.448395,0.448395);
  rgb(20pt)=(0.709441,0.460044,0.460044);
  rgb(21pt)=(0.726483,0.471405,0.471405);
  rgb(22pt)=(0.743134,0.482498,0.482498);
  rgb(23pt)=(0.759421,0.493342,0.493342);
  rgb(24pt)=(0.766356,0.517549,0.503953);
  rgb(25pt)=(0.773229,0.540674,0.514344);
  rgb(26pt)=(0.780042,0.562849,0.524531);
  rgb(27pt)=(0.786796,0.584183,0.534522);
  rgb(28pt)=(0.793492,0.604765,0.544331);
  rgb(29pt)=(0.800132,0.624669,0.553966);
  rgb(30pt)=(0.806718,0.643958,0.563436);
  rgb(31pt)=(0.81325,0.662687,0.57275);
  rgb(32pt)=(0.81973,0.6809,0.581914);
  rgb(33pt)=(0.82616,0.698638,0.590937);
  rgb(34pt)=(0.832539,0.715937,0.599824);
  rgb(35pt)=(0.83887,0.732828,0.608581);
  rgb(36pt)=(0.845154,0.749338,0.617213);
  rgb(37pt)=(0.851392,0.765493,0.625727);
  rgb(38pt)=(0.857584,0.781313,0.634126);
  rgb(39pt)=(0.863731,0.796819,0.642416);
  rgb(40pt)=(0.869835,0.812029,0.6506);
  rgb(41pt)=(0.875897,0.82696,0.658682);
  rgb(42pt)=(0.881917,0.841625,0.666667);
  rgb(43pt)=(0.887896,0.85604,0.674556);
  rgb(44pt)=(0.893835,0.870216,0.682355);
  rgb(45pt)=(0.899735,0.884164,0.690066);
  rgb(46pt)=(0.905597,0.897896,0.697691);
  rgb(47pt)=(0.911421,0.911421,0.705234);
  rgb(48pt)=(0.917208,0.917208,0.727166);
  rgb(49pt)=(0.922958,0.922958,0.748455);
  rgb(50pt)=(0.928673,0.928673,0.769156);
  rgb(51pt)=(0.934353,0.934353,0.789314);
  rgb(52pt)=(0.939999,0.939999,0.808969);
  rgb(53pt)=(0.945611,0.945611,0.828159);
  rgb(54pt)=(0.95119,0.95119,0.846913);
  rgb(55pt)=(0.956736,0.956736,0.865261);
  rgb(56pt)=(0.96225,0.96225,0.883229);
  rgb(57pt)=(0.967733,0.967733,0.900837);
  rgb(58pt)=(0.973185,0.973185,0.918109);
  rgb(59pt)=(0.978607,0.978607,0.935061);
  rgb(60pt)=(0.983999,0.983999,0.951711);
  rgb(61pt)=(0.989361,0.989361,0.968075);
  rgb(62pt)=(0.994695,0.994695,0.984167);
  rgb(63pt)=(1,1,1)
},
point meta max=1,
point meta min=0,
tick align=outside,
tick pos=left,
title={ANSYN},
xmajorgrids,
xlabel={$k$},
xmin=-0.5, xmax=24.5,
xtick style={color=black},
y dir=reverse,
ymajorgrids,
ylabel={$k^{\prime}$},
ymin=-0.5, ymax=24.5,
ytick style={color=black},
height=0.185\textwidth,
width=0.185\textwidth,
ticklabel style = {font=\footnotesize},
ylabel near ticks,
]
\addplot graphics [includegraphics cmd=\pgfimage,xmin=-0.5, xmax=24.5, ymin=24.5, ymax=-0.5] {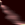};
\end{axis}

\end{tikzpicture}\vspace{-0.1cm}
\begin{tikzpicture}

\begin{axis}[
colorbar style={ylabel={}},
colormap={mymap}{[1pt]
  rgb(0pt)=(0.1178,0,0);
  rgb(1pt)=(0.195857,0.102869,0.102869);
  rgb(2pt)=(0.250661,0.145479,0.145479);
  rgb(3pt)=(0.295468,0.178174,0.178174);
  rgb(4pt)=(0.334324,0.205738,0.205738);
  rgb(5pt)=(0.369112,0.230022,0.230022);
  rgb(6pt)=(0.400892,0.251976,0.251976);
  rgb(7pt)=(0.430331,0.272166,0.272166);
  rgb(8pt)=(0.457882,0.290957,0.290957);
  rgb(9pt)=(0.483867,0.308607,0.308607);
  rgb(10pt)=(0.508525,0.3253,0.3253);
  rgb(11pt)=(0.532042,0.341178,0.341178);
  rgb(12pt)=(0.554563,0.356348,0.356348);
  rgb(13pt)=(0.576204,0.370899,0.370899);
  rgb(14pt)=(0.597061,0.3849,0.3849);
  rgb(15pt)=(0.617213,0.39841,0.39841);
  rgb(16pt)=(0.636729,0.411476,0.411476);
  rgb(17pt)=(0.655663,0.424139,0.424139);
  rgb(18pt)=(0.674066,0.436436,0.436436);
  rgb(19pt)=(0.69198,0.448395,0.448395);
  rgb(20pt)=(0.709441,0.460044,0.460044);
  rgb(21pt)=(0.726483,0.471405,0.471405);
  rgb(22pt)=(0.743134,0.482498,0.482498);
  rgb(23pt)=(0.759421,0.493342,0.493342);
  rgb(24pt)=(0.766356,0.517549,0.503953);
  rgb(25pt)=(0.773229,0.540674,0.514344);
  rgb(26pt)=(0.780042,0.562849,0.524531);
  rgb(27pt)=(0.786796,0.584183,0.534522);
  rgb(28pt)=(0.793492,0.604765,0.544331);
  rgb(29pt)=(0.800132,0.624669,0.553966);
  rgb(30pt)=(0.806718,0.643958,0.563436);
  rgb(31pt)=(0.81325,0.662687,0.57275);
  rgb(32pt)=(0.81973,0.6809,0.581914);
  rgb(33pt)=(0.82616,0.698638,0.590937);
  rgb(34pt)=(0.832539,0.715937,0.599824);
  rgb(35pt)=(0.83887,0.732828,0.608581);
  rgb(36pt)=(0.845154,0.749338,0.617213);
  rgb(37pt)=(0.851392,0.765493,0.625727);
  rgb(38pt)=(0.857584,0.781313,0.634126);
  rgb(39pt)=(0.863731,0.796819,0.642416);
  rgb(40pt)=(0.869835,0.812029,0.6506);
  rgb(41pt)=(0.875897,0.82696,0.658682);
  rgb(42pt)=(0.881917,0.841625,0.666667);
  rgb(43pt)=(0.887896,0.85604,0.674556);
  rgb(44pt)=(0.893835,0.870216,0.682355);
  rgb(45pt)=(0.899735,0.884164,0.690066);
  rgb(46pt)=(0.905597,0.897896,0.697691);
  rgb(47pt)=(0.911421,0.911421,0.705234);
  rgb(48pt)=(0.917208,0.917208,0.727166);
  rgb(49pt)=(0.922958,0.922958,0.748455);
  rgb(50pt)=(0.928673,0.928673,0.769156);
  rgb(51pt)=(0.934353,0.934353,0.789314);
  rgb(52pt)=(0.939999,0.939999,0.808969);
  rgb(53pt)=(0.945611,0.945611,0.828159);
  rgb(54pt)=(0.95119,0.95119,0.846913);
  rgb(55pt)=(0.956736,0.956736,0.865261);
  rgb(56pt)=(0.96225,0.96225,0.883229);
  rgb(57pt)=(0.967733,0.967733,0.900837);
  rgb(58pt)=(0.973185,0.973185,0.918109);
  rgb(59pt)=(0.978607,0.978607,0.935061);
  rgb(60pt)=(0.983999,0.983999,0.951711);
  rgb(61pt)=(0.989361,0.989361,0.968075);
  rgb(62pt)=(0.994695,0.994695,0.984167);
  rgb(63pt)=(1,1,1)
},
point meta max=1,
point meta min=0,
tick align=outside,
tick pos=left,
title={RESYN},
xmajorgrids,
xlabel={$k$},
xmin=-0.5, xmax=24.5,
xtick style={color=black},
y dir=reverse,
ymajorgrids,
ymin=-0.5, ymax=24.5,
ytick style={color=black},
height=0.185\textwidth,
width=0.185\textwidth,
yticklabels={,,},
ticklabel style = {font=\footnotesize}
]
\addplot graphics [includegraphics cmd=\pgfimage,xmin=-0.5, xmax=24.5, ymin=24.5, ymax=-0.5] {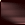};
\end{axis}

\end{tikzpicture}\vspace{-0.1cm}
\begin{tikzpicture}

\begin{axis}[
colorbar,
colorbar style={ylabel={}, width=0.25cm},
colormap={mymap}{[1pt]
  rgb(0pt)=(0.1178,0,0);
  rgb(1pt)=(0.195857,0.102869,0.102869);
  rgb(2pt)=(0.250661,0.145479,0.145479);
  rgb(3pt)=(0.295468,0.178174,0.178174);
  rgb(4pt)=(0.334324,0.205738,0.205738);
  rgb(5pt)=(0.369112,0.230022,0.230022);
  rgb(6pt)=(0.400892,0.251976,0.251976);
  rgb(7pt)=(0.430331,0.272166,0.272166);
  rgb(8pt)=(0.457882,0.290957,0.290957);
  rgb(9pt)=(0.483867,0.308607,0.308607);
  rgb(10pt)=(0.508525,0.3253,0.3253);
  rgb(11pt)=(0.532042,0.341178,0.341178);
  rgb(12pt)=(0.554563,0.356348,0.356348);
  rgb(13pt)=(0.576204,0.370899,0.370899);
  rgb(14pt)=(0.597061,0.3849,0.3849);
  rgb(15pt)=(0.617213,0.39841,0.39841);
  rgb(16pt)=(0.636729,0.411476,0.411476);
  rgb(17pt)=(0.655663,0.424139,0.424139);
  rgb(18pt)=(0.674066,0.436436,0.436436);
  rgb(19pt)=(0.69198,0.448395,0.448395);
  rgb(20pt)=(0.709441,0.460044,0.460044);
  rgb(21pt)=(0.726483,0.471405,0.471405);
  rgb(22pt)=(0.743134,0.482498,0.482498);
  rgb(23pt)=(0.759421,0.493342,0.493342);
  rgb(24pt)=(0.766356,0.517549,0.503953);
  rgb(25pt)=(0.773229,0.540674,0.514344);
  rgb(26pt)=(0.780042,0.562849,0.524531);
  rgb(27pt)=(0.786796,0.584183,0.534522);
  rgb(28pt)=(0.793492,0.604765,0.544331);
  rgb(29pt)=(0.800132,0.624669,0.553966);
  rgb(30pt)=(0.806718,0.643958,0.563436);
  rgb(31pt)=(0.81325,0.662687,0.57275);
  rgb(32pt)=(0.81973,0.6809,0.581914);
  rgb(33pt)=(0.82616,0.698638,0.590937);
  rgb(34pt)=(0.832539,0.715937,0.599824);
  rgb(35pt)=(0.83887,0.732828,0.608581);
  rgb(36pt)=(0.845154,0.749338,0.617213);
  rgb(37pt)=(0.851392,0.765493,0.625727);
  rgb(38pt)=(0.857584,0.781313,0.634126);
  rgb(39pt)=(0.863731,0.796819,0.642416);
  rgb(40pt)=(0.869835,0.812029,0.6506);
  rgb(41pt)=(0.875897,0.82696,0.658682);
  rgb(42pt)=(0.881917,0.841625,0.666667);
  rgb(43pt)=(0.887896,0.85604,0.674556);
  rgb(44pt)=(0.893835,0.870216,0.682355);
  rgb(45pt)=(0.899735,0.884164,0.690066);
  rgb(46pt)=(0.905597,0.897896,0.697691);
  rgb(47pt)=(0.911421,0.911421,0.705234);
  rgb(48pt)=(0.917208,0.917208,0.727166);
  rgb(49pt)=(0.922958,0.922958,0.748455);
  rgb(50pt)=(0.928673,0.928673,0.769156);
  rgb(51pt)=(0.934353,0.934353,0.789314);
  rgb(52pt)=(0.939999,0.939999,0.808969);
  rgb(53pt)=(0.945611,0.945611,0.828159);
  rgb(54pt)=(0.95119,0.95119,0.846913);
  rgb(55pt)=(0.956736,0.956736,0.865261);
  rgb(56pt)=(0.96225,0.96225,0.883229);
  rgb(57pt)=(0.967733,0.967733,0.900837);
  rgb(58pt)=(0.973185,0.973185,0.918109);
  rgb(59pt)=(0.978607,0.978607,0.935061);
  rgb(60pt)=(0.983999,0.983999,0.951711);
  rgb(61pt)=(0.989361,0.989361,0.968075);
  rgb(62pt)=(0.994695,0.994695,0.984167);
  rgb(63pt)=(1,1,1)
},
point meta max=1,
point meta min=0,
tick align=outside,
tick pos=left,
title={REAL},
xmajorgrids,
xlabel={$k$},
xmin=-0.5, xmax=24.5,
xtick style={color=black},
y dir=reverse,
ymajorgrids,
ymin=-0.5, ymax=24.5,
ytick style={color=black},
height=0.185\textwidth,
width=0.185\textwidth,
yticklabels={,,},
ticklabel style = {font=\footnotesize}
]
\addplot graphics [includegraphics cmd=\pgfimage,xmin=-0.5, xmax=24.5, ymin=24.5, ymax=-0.5] {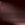};
\end{axis}

\end{tikzpicture}
	\caption{Attention maps produced by the \gls{ADRENALINE} model for exemplary chunks from all three evaluation datasets. The index \(k^{\prime}\) denotes the encoder time index, whereas \(k\) is the decoder time index.}
	\label{fig:attention_maps}
\end{figure}

Incorporating temporal information is crucial for effectively performing \gls{SEL}. Even though conventional \gls{GRU} or \gls{LSTM} architectures can capture long-term temporal dependencies, the attention-based framework utilized here seems to provide a better utilization of temporal information for predicting source activity and \glspl{DoA}. A possible reason for this is the ability of attention-based sequence-to-sequence models to incorporate multiple relevant time steps into the estimation, without taking into account their specific temporal order. This is not possible for (bidirectional) recurrent architectures, as they can only operate on the currently accumulated information in the forward and backward passes. Fig.~\ref{fig:attention_maps} depicts exemplary attention maps for all three evaluation corpora, which show that the attention weights have a larger spread in reverberant environments. This seems reasonable, as \gls{DoA} estimation under increased reverberation requires longer temporal contexts.

All methods yield comparable performance for estimating the correct number of sources. This is reflected by the frame recall metric shown in Tab.~\ref{tab:frame_recall}. The frame recall seems to degrade significantly in reverberant environments. This indicates that reverberation does not only provide a challenge for localization, but also hinders all investigated models in distinguishing between concurrent sound events.

\section{Conclusion and outlook}
\label{sec:conclusion}
This paper presents an attention-based sequence-to-sequence architecture for sound event localization. Compared to previous models based on standard recurrent neural network architectures, the proposed model better utilizes temporal information for estimating the activity and direction-of-arrival of sound events. An experimental evaluation on publicly available datasets for sound event localization supports the initial hypothesis that exploiting attentions for temporal focusing aids estimating the location of sound events. In addition, the proposed framework still provides many interesting opportunities to develop this approach further. By acknowledging the latest progress in natural language processing, a transformer-based architecture might also be applicable as an alternative to sequence-to-sequence models for sound event localization. Additionally, the incorporation of probabilistic estimates through, e.g., differentiable Bayesian techniques might further broaden the application range of the proposed framework.

\bibliographystyle{IEEEbib}
\bibliography{references}

\end{document}